\documentstyle[epsfig]{article}

\title{Applying Blind Chaos Control to Find Periodic Orbits}
\author{Daniel T. Kaplan\\
\em Department of Mathematics and Computer Science\\
\em Macalester College\\
\em St. Paul, Minnesota 55105}
\date{\today}

\newcommand{\figurewidth}{3.4in}

\begin{document}

\maketitle

\begin{abstract}
{\sf Abstract: Analysis of the PPF chaos control method used in biological
experiments shows that it can robustly control a wider class of
systems than previously believed, including those without stable
manifolds.  This can be exploited to
find the locations of unstable periodic orbits by varying the
parameters of the control system.}
\end{abstract}

\bigskip

One of the most surprising successes of chaos theory has been in biology: 
the experimentally demonstrated ability to control the timing of
spikes of electrical activity in complex and apparently chaotic
systems such as heart tissue \cite{Garfinkel} and brain tissue
\cite{Schiff}.  In these experiments, PPF control --- 
a modified formulation of OGY
control \cite{OGY} --- 
was applied to set the timing of external stimuli;
the controlled system showed stable periodic trajectories instead
of the irregular interspike intervals seen in the uncontrolled
system.  The mechanism of control in these experiments was interpreted
originally as analogous to that of OGY control: unstable periodic
orbits riddle the chaotic attractor and the electrical stimuli place
the system's state on the stable manifold of one of these periodic
orbits.

Alternative possible mechanisms for the experimental observations
have been described by Zeng and Glass \cite{zeng-glass} and Christini
and Collins \cite{collins}.  These authors point out that the
controlling external stimuli serve to truncate 
the interspike interval to
a maximum value.  When applied, the control stimulus sets the
next interval $s_{n+1}$ to be on the line
\begin{equation}
\label{eq:control-line}
s_{n+1} = {\cal A} s_{n} + {\cal C} .
\end{equation}  
We will call this relationship the ``control line.''
Zeng and Glass showed that
if the uncontrolled relationship between interspike intervals is a 
chaotic one-dimensional function, $s_{n+1} = f( s_{n} )$, then the
control system effectively flattens the top of this map and the controlled
dynamics may have fixed points or other periodic
orbits\cite{zeng-glass}.
Christini and Collins showed that behavior analogous to the
fixed-point control seen in the biological experiments
can be accomplished even in completely random systems\cite{collins}.  
Since neither
chaotic one-dimensional systems nor random systems have a stable
manifold, the interval-truncation interpretation of the biological experiments
is different than the OGY interpretation.  The interval-truncation method
differs also from OGY and related control methods in that the
perturbing control input is a fixed-size stimulus whose timing
can be treated as a continuous parameter.  This type of input is 
conventional in cardiology (e.g., \cite{hall1}).

In this Letter, we show that the state-trunction interpretation is
applicable in cases where there is a stable manifold of a
periodic orbit as well as in cases where there are only unstable manifolds.
We find that superior control can be achieved by
intentionally placing the system's state off of any stable manifold.
This suggests a powerful scheme for the rapid experimental
identification of fixed points and other periodic orbits in systems where 
interspike intervals are of interest.

The chaos control in \cite{Garfinkel} and \cite{Schiff} was implemented in two
stages.  First, interspike intervals $s_n$ from the uncontrolled,
``natural'' system were observed.  Modeling the system as a function
of two variables $s_{n+1} = f( s_n, s_{n-1} )$, the location
$s^\star$ of a putative unstable flip-saddle type fixed point 
and the corresponding stable eigenvalue 
$\lambda_s$ were
estimated from the data.\cite{so-sauer-ming}
(Since the fixed point is unstable, there is
also an unstable eigenvalue $\lambda_u$.)
The linear approximation to the stable
manifold lies on a line given by Eq. \ref{eq:control-line} with
${\cal A} = \lambda_s$ and ${\cal C} = (1 - \lambda_s) s^\star$.
Second, using estimated values of ${\cal A}$ and ${\cal C}$, the
control system was turned on.  Following each observed interval $s_n$,
the maximum allowed value of the next interspike interval was computed
as ${\cal S}_{n+1} = {\cal A} s_{n} + {\cal C}$.  
If the next interval naturally was shorter than ${\cal S}_{n+1}$ no
control stimulus was applied to the system.  Otherwise, an external
stimulus was provided to truncate the interspike interval 
at $s_{n+1} = {\cal S}_{n+1}$. \cite{footnote1}

In practice, the values of $s^\star$ and $\lambda_s$ for a real fixed
point of the natural system are known only
imperfectly from the data.  Insofar as the estimates are
inaccurate, the control system does not place the state on the true
stable manifold.  Therefore, we will analyze the controlled system
without presuming that ${\cal A}$ and ${\cal C}$ in
Eq. \ref{eq:control-line} correspond to the stable manifold.

If the natural dynamics of the system is modeled by
$s_{n+1} = f( s_n, s_{n-1} )$, the dynamics of the controlled system
is given by

\begin{equation}
\label{eq:control-dynamics}
s_{n+1} = \mbox{min} \left\{ 
\begin{array}{ll}
f( s_n, s_{n-1} ) & \mbox{Natural Dynamics} \\
{\cal A} s_n + {\cal C} & \mbox{Control Line}\\
\end{array}
\right.
\end{equation}

We can study the dynamics of the controlled system close to a natural
fixed point, $s^\star$, by approximating the natural dynamics linearly
\cite{linear-note}
as
\begin{eqnarray}
\label{eq:nat-lin-dynamics}
s_{n+1} = f(s_n, s_{n-1} ) & = &
(\lambda_s + \lambda_u ) s_n - \lambda_s \lambda_u s_{n-1} \nonumber \\
& & +
s^\star (1 + \lambda_s \lambda_u - \lambda_s - \lambda_u )
\end{eqnarray}
Since the controlled system (Eq. \ref{eq:control-dynamics}
is nonlinear even when $f()$ is linear,
it is difficult to analyze its behavior by algebraic iteration.  
Nonetheless, the controlled system can be studied in terms of
one-dimensional maps.  

Following any interspike interval
when the controlling stimulus has been applied, the system's state
$(s_n, s_{n-1})$ will lie somewhere on the control line.  From this
time onward the state will lie on an image of the control line even if
additional stimuli are applied during future interspike intervals.

\begin{figure}

\centerline{
\epsfig{file=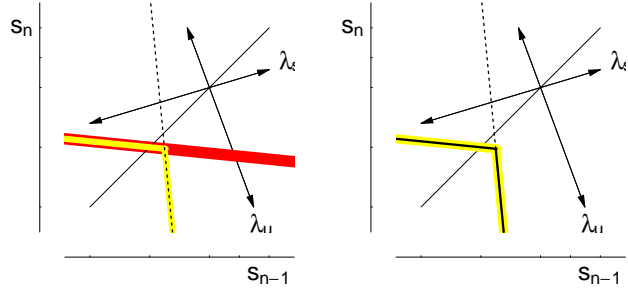,width=\figurewidth}
}
\medskip
\caption{
Successive images of the control line for a flip saddle with
$\lambda_s = 0.3$, $\lambda_u = -2.7$, ${\cal A} = -0.1$, and
$x^\star - s^\star = -.75$.  
Left: the control line is shown as
the broad dark gray line.  Its image under the natural dynamics is
shown as a dashed line; the image under the control dynamics is the
bent light gray line.  Right: the first image of the control line is 
replotted.  The image of this under the natural dynamics is the dashed line;
the image under the control dynamics is the solid line.
}
\end{figure}

Figure 1 (left) shows an example of how the dynamics result in a
simple one-dimensional map for the case where the natural dynamics
have a flip saddle ($\lambda_u < -1$ and $ 0 < \lambda_s < 1$) and
where the control line intersects the line of 
identity ($s_n = s_{n-1}$) below the natural fixed point $s^\star$.  The
stable and unstable manifolds are shown as arrows which intersect
at the location of the natural fixed point $s^\star$.  The control
line is shown as a broad dark gray stripe.  Its image under the
natural dynamics is shown as a thin dashed line.  At some points this
image is above the control line and is therefore truncated 
(in the vertical direction) by the control
stimulus to be on the control line.  Overall, the image of the control
line under the controlled dynamics is shown as the broad light gray
bent line.  In this case, the first, second, and all
successive images of the control line are
all the same: see Fig 1 (right).

Once the control stimulus has been applied,  the dynamics of the
controlled system are described by a one-dimensional map: 
the bent light gray line in Fig. 1.  The analysis of the
dynamics of this map is straightforward.  For the case shown in 
Fig. 1, the map has a fixed point (where the flat part of the map
intersects the line of identity).  Near this fixed point, the map is
identical to the control line, so the fixed point of the map is also
the ``controller fixed point,''
$$ x^\star = {\cal C}/(1-{\cal A})$$ 
where the control line intersects the line of identity.

\begin{figure}
\centerline{
\epsfig{file=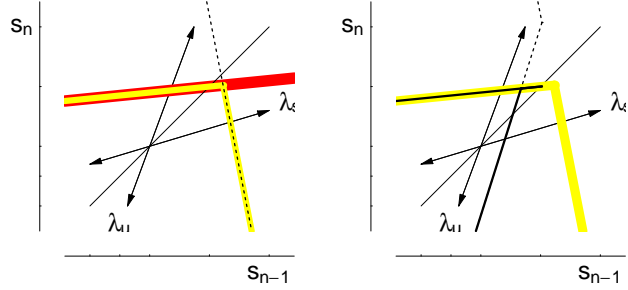,width=\figurewidth}
}
\medskip
\caption{
Successive images of the control line for a 
non-flip saddle: $\lambda_s = 0.3$, $\lambda_u = 2.7$, 
${\cal A} = -0.1$, $x^\star - s^\star = .5$.  
(See Fig. 1 for the key.)
The stable fixed point
of the controlled dynamics is located just to the right of the elbow
in the image of the control line.
}

\end{figure}

Figure 2 illustrates a more complicated case, a non-flip saddle with 
$x^\star > s^\star$, where successive images
of the control line do not all overlap.
In this case, successive images are not
identical, but there is still a stable fixed point of the controlled
dynamics at $x^\star$.  

The stability of the controlled dynamics fixed point 
and the size of its basin of attraction 
can be analyzed in terms of the control line and
its image.
When the previous interspike interval has been terminated by a control
stimulus, the state lies somewhere on the control line.  If the
controlled dynamics are to have a stable fixed point, this must be
at the controller fixed point $x^\star$ 
where the control line intersects the line of identity.
However, the controller fixed
point need not be a fixed point of the controlled dynamics.  For
example, if the image of the controller fixed point is below the
controller fixed point, then the interspike interval following a
stimulus will be terminated naturally.

For the controller fixed point to be a fixed point of the
controlled dynamics, we require that the natural 
image of the controller fixed
point be at or above the controller fixed point.  
One such situation,
for the flip saddle,
is illustrated in Fig. 1
where it can be seen \cite{cobweb-note}
that the natural image of a neighborhood of the control line near
$x^\star$ is above the
control line.  
Thus the
dynamics of the controlled system, close to $x^\star$, are given simply by
$$ s_{n+1} = {\cal A} s_n + {\cal C}$$
The fixed point of these dynamics is stable so long as 
$-1 < {\cal A} < 1$. 
In the case of a flip saddle, we therefore have a simple recipe for
successful state-truncation control: position $x^\star$
below the natural fixed point $s^\star$ and set 
$-1 < {\cal A} < 1$. 

Fixed points of the controlled dynamics can exist for natural
dynamics other than flip saddles.  This can be seen using the
following
reasoning: Let $\xi$ be the difference between the
controller fixed point and the natural fixed point:
$s^\star = x^\star + \xi$.  Then the natural image
of the controller fixed point can be found from
Eq. \ref{eq:nat-lin-dynamics} to be
\begin{eqnarray}
\label{eq:period-1-conditions}
s_{n+1} & = & (\lambda_s + \lambda_u) x^\star - 
\lambda_s \lambda_u x^\star \nonumber \\
& & + (1 + \lambda_s \lambda_u - \lambda_s - \lambda_u)( x^\star + \xi)
\end{eqnarray}
The condition that 
\begin{equation}
\label{eq:above-condition}
s_{n+1} \geq x^\star
\end{equation}
will be satisfied depending
only on $\lambda_s$, $\lambda_u$, and $\xi = s^\star - x^\star$.
In the case represented in Fig. 1, where 
$\xi < 0$ and $\lambda_u < -1$, the condition $s_{n+1} > x^\star$ will
be satisfied for any $\lambda_s < 1$.  This means that for any flip
saddle, so long as $x^\star < s^\star$, the point $x^\star$ will be a
fixed point of the controlled dynamics and will be stable so long
as $-1< {\cal A} < 1$.

Equations \ref{eq:period-1-conditions}  and 
\ref{eq:above-condition} imply that
control can lead to a stable fixed point for any type of fixed point except those for which
both $\lambda_u $ and $\lambda_s$ are greater than 1 
(so long as
$-1 < {\cal A} < 1$).  Since the
required relationship between $x^\star$ and $s^\star$ for a stable
fixed point of the controlled dynamics depends on the eigenvalues, it is
convenient to divide the fixed points into four classes, as given in
Table \ref{table:fpcases}

For example, for
the non-flip saddle shown in Fig. 2, the natural image of the control
line at $x^\star$ is above $x^\star$.  Thus, the controlled image will
be truncated (vertically) to be identical to $x^\star$ and therefore
the controller fixed point is also a fixed point of the controlled
dynamics.  This will be stable for $-1 < {\cal A} <1$, but with a
finite basin of attraction.

Beyond the issue of the stability of the fixed point of the controlled
dynamics, there is the question of the size of the fixed point's basin
of attraction.  Although the local stability of the fixed point is
guaranteed for the cases in Table \ref{table:fpcases} for 
$-1 < {\cal A} < 1$, the basin of attraction of this fixed point may
be small or large depending on ${\cal A}$, ${\cal C}$, $ s^\star$, 
$\lambda_u$ and $\lambda_s$.  For the case of Fig. 1, the basin is
finite when 
$|\lambda_s + \lambda_u - \lambda_s \lambda_u/{\cal A} | > 1 $.
In the case of Fig. 2, and for non-flip repellers generally, any
initial condition that is mapped to below the $\lambda_s$ eigenvector
will receed away from $x^\star$. \cite{upcoming}

\begin{table}

\centerline{\begin{tabular}{|l|c|c|c|}\hline
Type of FP & $\lambda_u$ & $\lambda_s$ & $x^\star$ Locat. \\ \hline \hline
Flip saddle & $\lambda_u < -1$ & $-1 < \lambda_s < 1$ & 
$ x^\star < s^\star $ \\ \hline
Saddle & $\lambda_u > 1$ & $-1 < \lambda_s < 1$ &
$ x^\star > s^\star $ \\ \hline
Single-flip repeller & $\lambda_u > 1$ & $\lambda_s < -1$ & 
$ x^\star > s^\star$ \\ \hline
Double-flip repeller & $\lambda_u < -1$ & $\lambda_s < -1$ &
$ x^\star < s^\star$ \\ \hline
Spiral (complex $\lambda$)& $ | \lambda_u |  > 1 $ & $|\lambda_s | > 1$ & $x^\star < s^\star$\\ \hline
\end{tabular}}

\caption{
Cases which lead to a stable fixed point for the controlled dynamics.
In all cases, it is assumed that $| {\cal A} | < 1$.
(For the cases where $ \lambda_s  < -1$, the subscript $s$ in
$\lambda_s$ is misleading in that the corresponding manifold is unstable.
For the spiral, there is no stable manifold.)
\label{table:fpcases}}

\end{table}

The endpoints of the basin of attraction can be derived 
analytically\cite{upcoming}.
The size of the basin of attraction will often be
zero when ${\cal A}$ and ${\cal C}$ are chosen to match the stable
manifold of the natural system.  Therefore, in order to make the basin large, 
it is advantageous intentionally to misplace the control line 
and to put $x^\star$ in the
direction indicated in Table \ref{table:fpcases}.  In addition,
control may be enhanced by setting ${\cal A} \neq \lambda_s$, for
instance
${\cal A} = 0$.
 
If the relationship between $x^\star$ and $s^\star$ is reversed from
that given in Table \ref{table:fpcases}, the controlled dynamics will
not have a stable fixed points.  To some extent, these can also be
studied using one-dimensional maps.
The flip saddle and double-flip
repeller can display stable period-2 orbits and chaos.  For the non-flip
saddle and single-flip repeller, control is unstable 
when $x^\star < s^\star$.

The fact that control may be successful or even enhanced when 
${\cal A}$ and ${\cal C}$ are not matched to $\lambda_s$ and
$s^\star$ suggests that it may be useful to reverse the experimental procedure
often followed in chaos control.  Rather than first identifying the
parameters of the natural unstable fixed points and then applying the
control, one can blindly attempt control and then deduce the natural
dynamics from the behavior of the controlled system.  This use of PPF
control is reminiscent of pioneering studies that used periodic
stimulation to demonstrate the complex dynamics of biological
preparations\cite{guevara}.

As an example, consider the Henon map:
$$ s_{n+1} = 1.4 + 0.3 s_{n-1} - s_{n}^2 $$
This system has two distinct fixed points.  There is a flip-saddle at
$s^\star = 0.884$ with $\lambda_u = -1.924$ and $\lambda_s = 0.156$
and a non-flip saddle at $s^\star = -1.584$ with 
$\lambda_u = 3.26 $ and $\lambda_s = -0.092$.  In addition, there is
an unstable flip-saddle orbit of period 2 following the sequence 
$1.366 \rightarrow -0.666 \rightarrow 1.366$.  There are no real
orbits of period 3, but there is an unstable orbit of period 4
following the sequence $.893 \rightarrow .305 \rightarrow 1.575
\rightarrow -.989 \rightarrow .893 $.  These facts can be deduced by
algebraic analysis of the equations.  

\begin{figure}
\centerline{
\epsfig{file=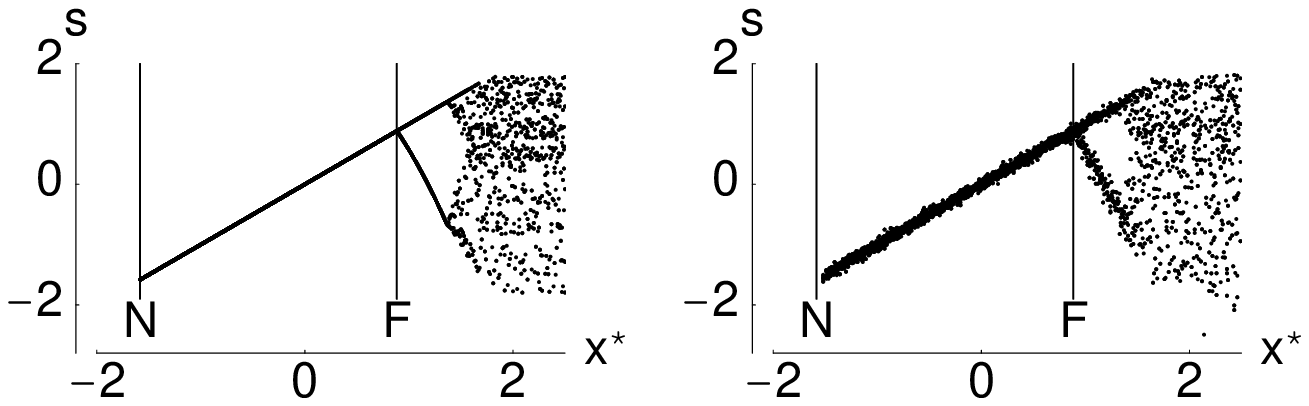,width=\figurewidth}
}
\medskip
\begin{caption}{
Bifurcation diagrams for the truncation controlled Henon map.
Left: No noise.  Right: Dynamical noise added (N(0,0.05)).
}
\end{caption}

\end{figure}

In an experiment using the controlled system, the control parameter
$x^\star = {\cal C}/(1-{\cal A})$ can be varied.  The theory presented
above indicates that the controlled system should undergo a
bifurcation as $x^\star$ passes through $s^\star$.  Figure 3 shows the
bifurcation diagram for the controlled Henon system 
(with ${\cal A}= 0$).  For each value of $x^\star$, the controlled
system was iterated from a random initial condition and the values of
$s_n$ plotted after allowing a transient to decay\cite{footnote2}.
A bifurcation
from a stable fixed point to a stable period 2 as $x^\star$ passes
through the flip-saddle value of $s^\star = 0.884$.  A different type
bifurcation occurs at the non-flip saddle 
fixed point at $s^\star = -1.584$.  To the left of the bifurcation
point, the iterates are diverging to $- \infty$ and are not plotted.

Adding gaussian dynamical noise (of standard deviation $0.05$) does
not substantially alter the bifurcation diagram, suggesting
that examination of the truncation control bifurcation diagram 
may be a practical way to read off the location of the unstable fixed
points in an experimental preparation.

\begin{figure}
\centerline{
\epsfig{file=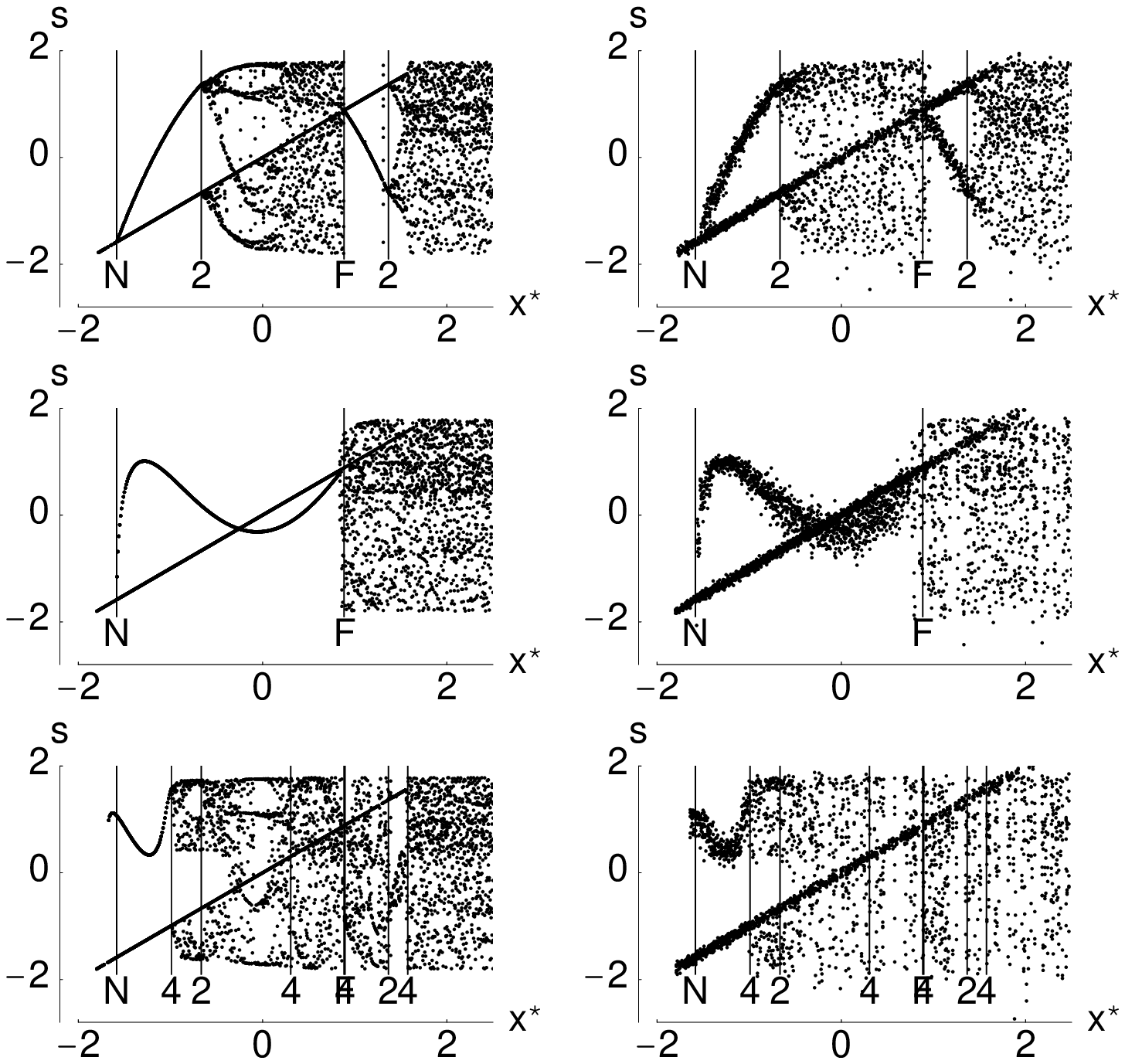,width=\figurewidth}}
\medskip
\begin{caption}{
Bifurcation diagrams for the truncation controlled Henon map where
truncation control is activated only every second (top), third (middle), or
fourth (bottom) iteration.
Left: No noise.  Right: Dynamical noise added (N(0,0.05)).
The locations of the true fixed points are marked with vertical lines:
F flip-saddle; N non-flip saddle; 2 period-2 orbit; 4 period-4 orbit.
}
\end{caption}

\end{figure}

By activating the truncation control after every second, third or
fourth iteration, it is possible to find periodic orbits of period 2,
3, and 4 respectively.  The bifurcation diagrams are shown in Fig. 4.
The location of the period-2 orbits can be clearly discerned 
even in the
presence of noise.  No period-3 orbit is indicated.
Noise obscures the location of all but one of the
period-4 points.

Unstable periodic orbits can be difficult to find in uncontrolled
dynamics because thre is typically little data near such orbits.
Application of PPF control, even blindly, can stabilize such orbits
and dramatically improve the ability to locate them.  This, and the
robustness of the control, may prove
particularly useful in biological experiments where orbits may drift
in time as the properties of the system change. \cite{gauthier}

We would like to acknowledge helpful conversations with Thomas
Schreiber and Leon Glass.


\begin{thebibliography}{100}
\bibitem{footnote1}
This truncation is reminiscent of
the mythical Greek inkeeper Procrustes.  Travellers too tall for his
bed were truncated to fit in a rather gruesome manner.  Although PPF
originally was intended to stand for ``Proportional Perturbation
Feedback,'' the nonlinear all-or-nothing 
nature of the perturbation suggests that an
alternative expansion is more appropriate: ``Procrustean Perturbation 
Feedback.''
\bibitem{linear-note}
Eq. \ref{eq:nat-lin-dynamics} is simply the linear equation
$s_{n+1} = a s_n + b s_{n-1} + c$ with $a$, $b$, and $c$ set to give
eigenvalues $\lambda_s$ and $\lambda_u$ and fixed point $s^\star$.
\bibitem{cobweb-note}
The image of a
single point on the control line can be found graphically using the
cobweb method.  From the point, trace horizontally to the line of
identity, then vertically to the image of the control line.  If
following the natural dynamics, trace to the natural image.  If
following the controlled dynamics, trace toward the natural image but
stop at the control line if that is below the natural image.
\bibitem{upcoming} 
Details will be provided in a subsequent communication.
\bibitem{footnote2} No attempt
was made to wait until the system state naturally approached a
putative fixed point.  The truncation control was active at all time.
The decay of the transient typically took only a few iterations.
\end{thebibliography}
\end{document}